\documentstyle[twocolumn]{venice97}
\input{psfig.tex}

\begin{document}

\setlength{\parindent}{0pt}
\setlength{\parskip}{ 10pt plus 1pt minus 1pt}
\setlength{\hoffset}{-1.5truecm}
\setlength{\textwidth}{ 17.1truecm }
\setlength{\columnsep}{1truecm }
\setlength{\columnseprule}{0pt}
\setlength{\headheight}{12pt}
\setlength{\headsep}{20pt}
\pagestyle{veniceheadings}

\title{\bf METAL ABUNDANCES OF ONE HUNDRED HIPPARCOS DWARFS }

\author{{\bf R.G.~Gratton$^1$, E.~Carretta$^2$, G.~Clementini$^2$,
C. Sneden$^3$} \vspace{2mm} \\
$^1$Osservatorio Astronomico di Padova, Vicolo dell'Osservatorio 5, 35122
  Padova, ITALY\\
$^2$Osservatorio Astronomico di Bologna, ITALY\\
$^3$Department of Astronomy, The University of Texas at Austin }

\maketitle

\begin{abstract}

Abundances for Fe, O, and the $\alpha-$elements (Mg, Si, Ca, and Ti) have been
derived from high resolution spectra of a sample of about one hundred dwarfs
with high precision parallaxes measured by HIPPARCOS. The stars have metal
abundances in the range $-2.5<$[Fe/H]$<0.2$. The observational data set
consists of high dispersion ($20,000<R<70,000$), high $S/N$\ ($>200$) spectra
collected at the Asiago and McDonald Observatories. The abundance analysis
followed the same precepts used by Gratton et al. (1997a) for $\sim 300$~field
stars and for giants in 24 globular clusters (\cite{cg97}), and includes
corrections for departures from LTE in the formation of O lines. Our main
results are:
 
1. the equilibrium of ionization of Fe is well satisfied in late F -- early
K dwarfs 

2. O and $\alpha-$elements are overabundant by $\sim 0.3$~dex

This large homogeneous data set was used in the derivation of accurate ages for
globular clusters (See paper by Gratton et al. at this same Meeting). 
\vspace {5pt} \\


Key~words: Stars: chemical abundances - Stars: basic parameters

\end{abstract}

\section{INTRODUCTION}

HIPPARCOS has provided parallaxes with accuracies of $\sim$ 1 mas for several
hundreds dwarfs. We had access to data for about 100 dwarfs with metal
abundances in the range $-2.5<$[Fe/H]$<0.2$\ and have used them in a thorough
revision of the ages of the oldest globular clusters derived by Main Sequence
(MS) fitting technique. A crucial step in the derivation of ages via this
method is the assumption that the nearby subdwarfs have the same chemical
composition of the globular cluster main sequence stars. This assumption was
verified through a careful abundance analysis of the vast majority of nearby
dwarfs with HIPPARCOS parallaxes available to us. 

Our data set and the HIPPARCOS parallaxes were also used to test whether an
appreciable Fe overionization occurred in the atmosphere of late F -- early K
dwarfs (\cite{b90}; \cite{mz96}). This was done by comparing abundances
provided by neutral and singly ionized lines, once the surface gravity of each
program star had be derived from its mass, temperature and luminosity rather
then from the equilibrium of ionization of Fe. 

Finally, our abundances are fully consistent with those presented by Gratton et
al. (1997a) for about 300 field dwarfs. A large, homogenous data base of high
accuracy (errors $\sim 0.07$~dex) abundances computed with the Kurucz (1993)
model atmospheres is now available and can be used to recalibrate photometric
and low $S/N$\ spectroscopic abundances. 

\section{BASIC DATA FOR SUBDWARFS}

Average V magnitudes and colors (Johnson $B-V$\ and $V-K$, and Str\"omgren
$b-y$, $m_1$\ and $c_1$) for the programme stars were obtained from a careful
discussion of the literature data. We used also the Tycho $V$\ magnitudes and
$B-V$\ colors, after correcting them for the very small systematic difference
with ground-based data. 

Absolute magnitudes $M_V$ were derived combining apparent V magnitudes and
Hipparcos parallaxes. No Lutz-Kelker corrections were applied. Lutz-Kelker
corrections (\cite{lk73}) take into account that stars with parallaxes measured
too high are more likely to be included in a sample if the sample selection
criteria are based on the parallaxes themselves. Since our sample was selected
before the HIPPARCOS parallaxes were known; Lutz-Kelker corrections should not
be applied when the whole sample is considered, as we do when comparing the
abundances obtained from Fe~I and Fe~II lines. 

Multiple high precision radial velocity observations exist for a large fraction
of our objects (80 out of 99). Twenty stars in the sample are known and four
are suspected spectroscopic binaries. Two further stars display very broad
lines in our spectra, possibly due to fast rotation. They were discarded. A few
other stars display some IR excess, which also may be a signature of binarity.
No evidence for binarity disturbing the present analysis exists for the
remaining stars. 

Sixty-eight out of the 99 stars of our sample are included in Carney et al.
(1994) catalogue. Reddening estimates are given for 58 of them. All but two
have zero values. We have thus assumed a zero reddening for all the programme
stars. 

\section{OBSERVATIONS AND REDUCTIONS}

High dispersion spectra for about two thirds of the programme stars were
acquired using the 2D-coud\`e spectrograph of the 2.7~m telescope at McDonald
Observatory and the REOSC echelle spectrograph at the 1.8~m telescope at Cima
Ekar (Asiago). McDonald spectra have a resolution $R=70,000$, $S/N\sim 200$,
and spectral coverage from about 4,000 to 9,000 \AA; they are available for 21
stars (most with [Fe/H]$<-0.8$). Cima Ekar telescope provided spectra with
resolution $R=15,000$, $S/N\sim 200$, and two spectral ranges (4,500$< \lambda
<$7,000 and 5,500 $<\lambda<$8,000~\AA) for 65 stars. 

Equivalent widths $EW$s of the lines were measured by means of a gaussian
fitting routine applied to the core of the lines; appropriate average
corrections were included to take into account the contribution of the damping
wings. Only lines with $\log {EW/\lambda}<-4.7$\ were used in the final
analysis (corrections to the $EW$s for these lines are $\leq 7$~m\AA, that is
well below 10~per cent). The large overlap between the two samples (14 stars)
allowed us to tie the Asiago $EW$s to the McDonald ones. 

External checks on our $EW$s are possible with Edvardsson et al (1993:
hereinafter E93) and Tomkin et al. (1992: hereinafter TLLS). Comparisons
performed using McDonald $EW$s alone show that they have errors of $\pm
4$~m\AA. From the r.m.s. scatter, $\sigma$, between Asiago and McDonald $EW$s,
we estimate that the former have errors of $\pm 6.7$~m\AA. When Asiago and
McDonald $EW$s are considered together, we find average residuals (us-others)
of $-0.2\pm 1.0$~m\AA\ (39 lines, $\sigma=6.1$~m\AA) and $+0.8\pm 1.0$~m\AA\
(36 lines, $\sigma=5.9$~m\AA) with E93 and TLLS, respectively. 

\section{ANALYSIS}

\subsection{Atmospheric Parameters}

The abundance derivation followed precepts very similar to the reanalysis of
$\sim 300$\ field and $\sim 150$\ globular cluster stars described in Gratton
et al. (1997a) and Carretta \& Gratton (1997). The same line parameters were
adopted. The effective temperatures were derived from $B-V$, $b-y$, and $V-K$\
colours using the iterative procedure outlined in Gratton et al. (1997a). 
Atmospheric parameters are derived as follows : 
\begin{enumerate}
\item we assume as input values $\log g=4.5$\ and the metal abundance derived
from the $uvby$\ photometry using the calibration of Schuster \& Nissen (1989)
\item $T_{\rm eff}$\ is then derived from the colours, using the empirical
calibration of Gratton et al. (1997a) for population I stars (assumed to be
valid for [Fe/H]=0), and the abundance dependence given by Kurucz (1993) models
\item a first iteration value of $\log g$\ is then derived from the absolute
bolometric magnitude (derived from the apparent $V$\ magnitude, parallaxes from
Hipparcos, and bolometric corrections $BC$\ from Kurucz 1993), and masses
obtained by interpolation in $T_{\rm eff}$\ and [A/H] within the Bertelli et
al. (1997) isochrones 
\item steps 2 and 3 are iterated until a consistent set of values is
obtained for $T_{\rm eff}$, $\log g$, and [A/H] 
\item the $EW$s are then analyzed, providing new values for $v_{\rm t}$\ and 
[A/H] (assumed to be equal to [Fe/H] obtained from neutral lines) 
\item the procedure is iterated until a new consistent set of parameters
is obtained
\end{enumerate}

\subsection{Error analysis}

Random errors in $T_{\rm eff}$\ ($\pm 45$~K) were  obtained by comparing
temperatures derived from different colours. Systematic errors may be larger;
the $T_{\rm eff}$-scale used in this paper is discussed in detail in Gratton et
al. (1997a). We assume that systematic errors in the adopted $T_{\rm eff}$'s
are $\leq 100$~K. 

Random errors in the gravities ($\pm 0.09$~dex) are estimated from the errors
in the masses (1.2~per~cent), $M_V$'s (0.18 mag), and in the $T_{\rm eff}$'s
(0.8~per~cent), neglecting the small contribution due to $BC$'s. Systematic
errors ($\pm 0.04$~dex) are mainly due to errors in the $T_{\rm eff}$\ scale
and in the solar $M_V$\ value. 

Random errors in the microturbulent velocities can be estimated from the
residuals around the fitting relation in $T_{\rm eff}$\ and $\log g$. We obtain
values of 0.47 and 0.17~km~s$^{-1}$\ for the Asiago and McDonald spectra,
respectively. 

Random errors in the $EW$s and the line parameters significantly affect the
abundances when few lines are measured for a given specie. Errors should scale
as $\sigma/\sqrt{n}$\, where $\sigma$\ is the typical error in the abundance
from a single line (0.14~dex for the Asiago spectra, and 0.11 dex for the
McDonald ones) and $n$\ is the number of lines used in the analysis. However,
errors may be larger if all lines for a given element are in a small spectral
range. Furthermore, undetected blends may contribute significantly to errors
when the spectra are very crowded (mainly Asiago spectra of cool, metal-rich
stars). 

Random errors in the model metal abundance are obtained by summing up
quadratically the errors due to the other sources. Systematic errors can be due
to non-solar abundance ratios. They can be as large as $\sim 0.2$~dex in the
metal-poor stars ([Fe/H]$<-0.5$), where O and the other $\alpha-$elements are
overabundant by $\sim 0.3$~dex. 

Random errors in the Fe abundances are $\sim 0.07$ and $\sim 0.04$~dex for 
abundances derived from Asiago and McDonald spectra, respectively. Systematic 
errors ($\sim 0.08$~dex) are mainly due to the $T_{\rm eff}$\ scale.

\subsection{Comparison with other abundances}

On average, differences (Asiago$-$McDonald) in the Fe abundances are $-0.01\pm
0.02$~dex (12 stars, $\sigma=0.07$~dex). Analogous differences for the [O/Fe]
and [$\alpha$/Fe] ratios are $+0.02\pm 0.08$~dex (5 stars, $\sigma=0.17$~dex),
and $+0.01\pm 0.03$~dex (12 stars, $\sigma=0.10$~dex). 

E93 measured abundances for $\sim 200$~dwarfs; six stars are in common with our
sample. Abundance residuals (our analysis$-$E93) are $+0.08\pm 0.03$, $-0.02\pm
0.03$, and $+0.02\pm 0.02$~dex for [Fe/H], [O/Fe], and [$\alpha$/Fe],
respectively. Residual differences are mainly due to our use of a higher
temperature scale (our $T_{\rm eff}$'s are larger by $63\pm 12$~K). We have six
stars in common with TLLS, which used a restricted wavelength range. Average
differences (ours$-$TLLS) are: $+0.34\pm 0.04$\ and $-0.31\pm 0.07$~dex for
[Fe/H] and [O/Fe], respectively. They are due to different assumption in the
analysis: (i) our temperature scale is higher; (ii) TLLS used a different solar
model; (iii) our non-LTE corrections to the O abundances are slightly larger.
Finally, Gratton et al. (1997a) made a homogenous reanalysis of the original
$EW$s for $\sim 300$~metal-poor field stars. On average, the present Fe
abundances are larger by $0.02\pm 0.02$~dex (11 stars, $\sigma=0.06$~dex).
Since the same analysis procedure is adopted, these differences are entirely
due to random errors in the $EW$s and in the adopted colours. In the following,
we assume that Gratton et al. abundances are on the same scale of the present
analysis. 

\subsection{Fe abundances}

Since gravities are derived from masses and luminosities rather than from the
equilibrium of ionization for Fe, we may test if predictions based on LTE are
satisfied for the program stars.

\begin{figure}
  \begin{center}
    \leavevmode
\centerline{\psfig{figure=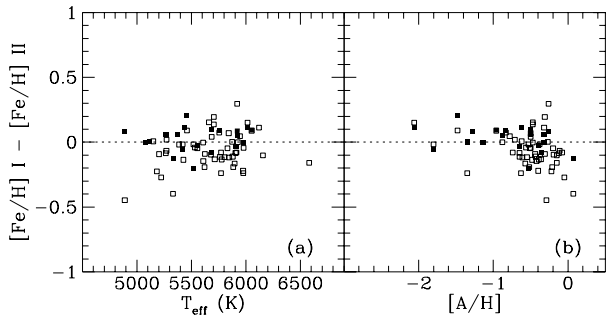,width=8.0cm}}
  \end{center}
  \caption{\em Run of the difference between the abundances derived from neutral and
singly ionized Fe lines as a function of temperature (panel $a$) and overall
metal abundance (panel $b$). Open squares are abundances obtained from the Asiago
spectra; filled squares are abundances obtained from the McDonald spectra }
\label{f:fe1fe2} 
\end{figure}

In Figure~\ref{f:fe1fe2} we plot the difference between abundances of Fe
obtained from neutral and singly ionized lines against effective temperature
and metal abundance. Different symbols refer to results obtained from McDonald
and Asiago spectra, respectively. McDonald spectra have a higher weight because
the higher resolution allowed us to measure a larger number of Fe~II lines
($10\sim 20$), and errors in the $EW$s are smaller; very few Fe~II lines could
be measured in the crowded spectra of cool and/or metal-rich stars observed
from Asiago. Average differences between abundances given by Fe~I and II lines
are $0.025\pm 0.020$ (21~stars, $\sigma=0.093$~dex) for the Mc Donald spectra,
and $-0.063\pm 0.019$ (52~stars, $\sigma=0.140$~dex) for the Asiago spectra.
The scatter obtained for McDonald spectra agrees quite well with the expected
random error of 0.085~dex. The average value is consistent with LTE if the
adopted $T_{\rm eff}$\ scale is too high by $\sim 20$~K, well within the quoted
error bar of $\pm 100$~K. The lower mean difference obtained for the Asiago
spectra is due to a few cool metal-rich stars which have very crowded spectra.
Very few Fe~II lines could be measured in these spectra and the line-to-line
comparison  with the superior McDonald data suggests that even these lines may
be affected by blends. 

We conclude that {\bf the equilibrium of ionization for Fe is well satisfied in
the late F -- K dwarfs of any metallicity in our sample}. This result depends
on the adopted temperature scale. 

Our empirical result agrees very well with the extensive statistical
equilibrium calculations for Fe by Gratton et al. (1997b). In that paper, the
uncertain collisional cross sections were normalized in order to reproduce the
observations of the RR Lyraes, where overionization is expected to be much
larger than in late F -- K dwarfs. The lower limit to collisional cross
sections given by the absence of detectable overionization in RR Lyrae spectra
(\cite{c95}) implies that LTE is a very good approximation for the formation of
Fe lines in dwarfs. 

\begin{figure}
  \begin{center}
   \leavevmode
\centerline{\psfig{figure=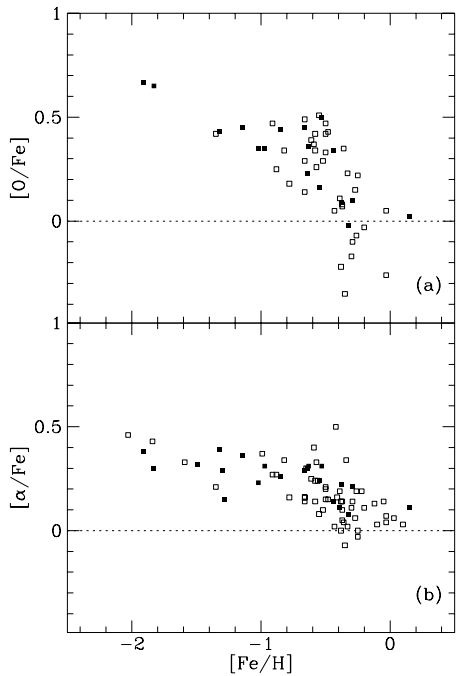,width=6.0cm}}
  \end{center}
  \caption{\em Runs of the overabundances of O (panel $a$) and $\alpha-$elements
(panel $b$) as a function of [Fe/H] for the programme subdwarfs. Filled
squares are abundances from McDonald spectra; open squares are abundances
from Asiago spectra }
\label{f:fig1} 
\end{figure}

\subsection{O and $\alpha-$element abundances}

O abundances were derived from the permitted IR triplet, and include non-LTE
corrections computed for each line in each star following the precepts of
Gratton et al. (1997b). We find that O and the other $\alpha-$elements are
overabundant in stars with [Fe/H]$<-0.5$ (see Figure~\ref{f:fig1}): 
$$ {\rm [O/Fe]}= 0.38\pm 0.13$$
$$ {\rm [\alpha/Fe]}= 0.26\pm 0.08,$$
(error bars are the r.m.s. scatter of individual values around the mean). The
moderate O excess derived from the IR permitted lines is a consequence of the
rather high temperature scale adopted. When this adoption is made, abundances
from permitted OI lines agree with those determined from the forbidden [OI] and
the OH lines. 

The present abundances agree very well with those derived in Gratton et al.
(1997c). Note also that the overabundance of O and $\alpha-$ elements found for
the field subdwarfs is similar to the excesses found for globular cluster
giants (apart from those stars affected by the O-Na anticorrelation, see
Kraft 1994).

\section{CALIBRATION OF PHOTOMETRIC ABUNDANCES}

Once combined with the abundances obtained by Gratton et al. (1997a), the
sample of late F to early K-type field stars with homogenous and accurate high
dispersion abundances adds up to nearly 400 stars. Schuster \& Nissen (1989)
have shown that rather accurate metal abundances for late F to early K-type can
be obtained using Str\"omgren $uvby$\ photometry (available for a considerable
fraction of the HIPPARCOS stars). Furthermore, the extensive binary search by
Carney et al. (1994) has provided a large number of metal abundances derived
from an empirical calibration of the cross correlation dips for metal-poor
dwarfs. 

\begin{figure}
  \begin{center}
    \leavevmode
\centerline{\psfig{figure=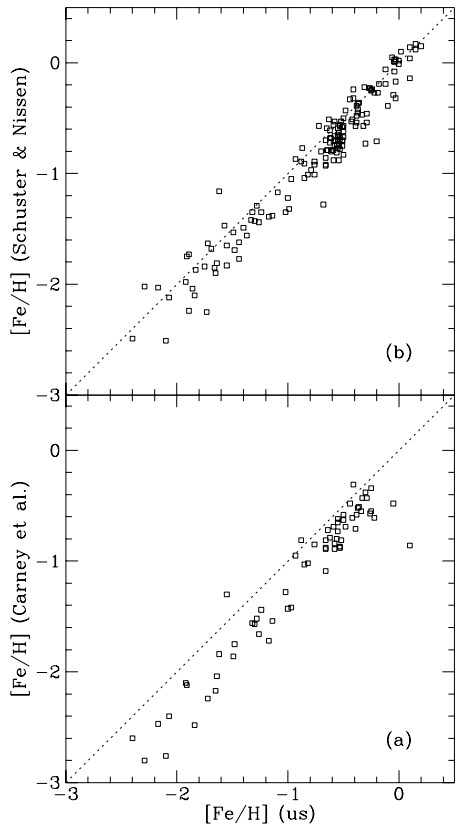,width=6.0cm}}
  \end{center}
  \caption{\em Comparison between the abundances obtained from high dispersion
spectra (present analysis or Gratton et al. 1997), and those provided by the
original calibration of Schuster \& Nissen (1989, panel $a$) and 
Carney et al. (1994, panel $b$) }
\label{f:cal} 
\end{figure}

We have recalibrated these abundance scales. Schuster \& Nissen (1989)
abundances only differs for a zero-point offset (see panel $a$\ of
Figure~\ref{f:cal}); the mean difference is: 
\begin{equation}
{\rm [Fe/H]}_{\rm us}={\rm [Fe/H]}_{\rm SN} + (0.102\pm 0.012),
\end{equation}
based on 152 stars (the r.m.s. scatter for a single star is 0.151~dex).

In the case of Carney et~al. (1994, panel $b$\ of Figure~\ref{f:cal}), a small
linear term is also required. The best fit line (66 stars) is: 
\begin{equation}
{\rm [Fe/H]}_{\rm us}=(0.94\pm 0.03){\rm [Fe/H]}_{\rm C94} + (0.18\pm 0.17),
\end{equation}

The offsets between the high dispersion abundances and those provided by
\cite*{sn89} and \cite*{c94} are mainly due to different assumptions about the
solar abundances in the high dispersion analyses originally used in the
calibrations of \cite*{sn89} and \cite*{c94}.

\end{document}